 \documentclass[final,3p,times,twocolumn]{elsarticle}
\usepackage{amssymb}
\usepackage{amsmath}
\usepackage{tikz}
\usepackage{lipsum}
\PassOptionsToPackage{square,comma,numbers,sort&compress}{natbib}
\usepackage{url} %
\usepackage{gensymb}
\journal{Nuclear Physics A}
\begin{document}

\begin{frontmatter}

\title{Measurement of Radon-222 concentration in N$_{2}$ using an activated charcoal trap}

\author[1]{N. Fatemighomi}
\author[1]{Y. Ahmed}
\author[1]{S.\,M.\,A.\,Hussain}
\author[1]{J. Lu}
\author[1]{A. Pearson}
\author[1]{J. Suys}
\affiliation[1]{organization={SNOLAB},
       addressline={Creighton Mine \#9},
             city={Sudubury},
             postcode={P3Y 1N2},
             state={ON},
             country={Canada}}

\begin{abstract}
Radon-222 is a limiting background in many leading dark matter and low energy neutrino experiments. One way to mitigate Radon-222 is to fill external experimental components with a clean cover gas such as N$_{2}$. At the SNOLAB facility in Canada, the $^{222}$Rn concentration in the cover gas systems of the experiments are monitored  using a radon assay board developed by the SNO collaboration.    To improve the sensitivity of N$_{2}$ assays, a new trapping mechanism  based on activated charcoal has been developed.  The trap was purified and tested at SNOLAB. The methods for determining the efficiency, background, and sensitivity of the trap were described. Additionally, as part of the efficiency measurement, a radon calibration source was developed and characterized.  
\end{abstract}

\end{frontmatter}

\section{Introduction}
\label{introduction}
\begin{figure}[h]
    \centering
    \includegraphics[width=1.0\linewidth]{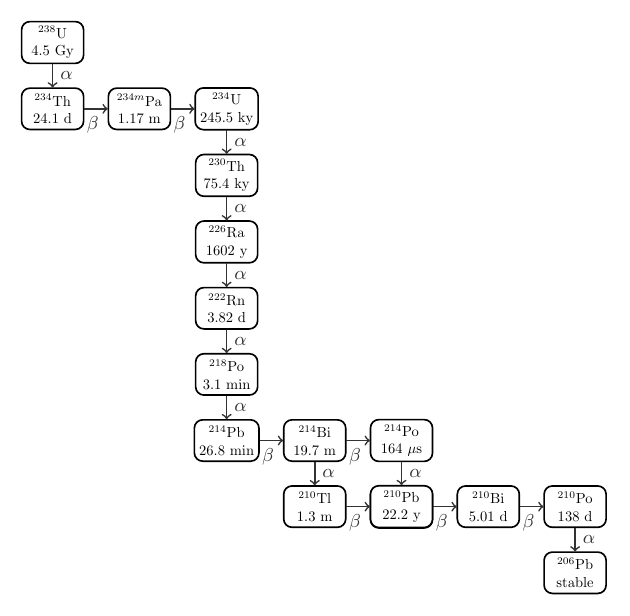}
    \caption{$^{238}$U decay chain.}
    \label{fig-u238}
\end{figure}

The purpose of this work is to introduce a method for measuring Radon-222 ($^{222}$Rn) concentration in the nitrogen (N$_{2}$) cover gas systems of experiments hosted at SNOLAB,  an underground science facility located 2 km deep in the Vale Creighton mine near Sudbury, Ontario. 
Experiments such as SNO+ \cite{snoplus}, DEAP-3600 \cite{deap3600}, and   high-purity germanium detectors~\cite{ug-radonlevel} use boil-off N$_{2}$ from SNOLAB's liquid nitrogen (LN) plants~\cite{LN-plant} and Grade 5 N$_{2}$ cylinder cylinders~\cite{n2} to control the background induced by $^{222}$Rn progeny. %
Radon-222 progeny can produce some of the most troublesome radioactive backgrounds for these experiments.  $^{222}$Rn is a radioactive noble gas with  a half-life of  3.8 days. It is the progeny of $^{226}$Ra in  the $^{238}$U decay chain as shown in Figure~\ref{fig-u238}.  $^{222}$Rn can enter detectors through leaks, surface emanation, and diffusion. %
Accurate measurement of the radon concentration in the cover gas volumes of experiments is crucial, as it helps  model their external radon-induced backgrounds. 

Radon-222 assay refers to the measurement of the concentration  of $^{222}$Rn in a given gas or water sample as well as the radon emanation and diffusion rate from a material surface. The SNO collaboration developed a technique to perform material and ultra pure water radon assays~\cite{LIU1993291}\cite{waterpaper}. In this technique,  radon atoms are extracted from a sample material or from water droplets, then trapped  in a radon trap which is at -196~$\degree$C. The radon trap in these two applications  is filled with bronze wool and is kept under vacuum.   The bronze wool is used to stop the radon atoms by increasing the surface area, while allowing the other gasses such as N$_{2}$ and O$_{2}$ to pass through.  The gas flow rate for these applications is $< 0.1$~ standard litter per minute (SLPM). 
This technique with the bronze wool as radon trapping mechanism  initially  adopted by the authors to perform N$_{2}$ gas assays, in which the gas flow rate is $\ge$~1~SLPM.    However, it was observed that the radon trapping efficiency decreases with the  duration that gas flows through the trap\cite{Adilthesis}. This limits the maximum volume of N$_{2}$ that can be sampled with this technique and, consequently, restricts the assay sensitivity.  We hypothesize that this is due to the increasing temperature of the trap as more N$_{2}$  passes through, which leads to the  releasing of the radon atoms. This observation motivated us to develop and add a new radon trap filled with a material which has higher heat capacity compared to  bronze wool and use it as a pre-trap to the existing system. 

Activated charcoal has at least twice the heat capacity of bronze wool~\cite{heatcapacity}. Furthermore, activated coconut charcoal, with its high adsorption capabilities, is  highly effective at trapping radon atoms and is commonly used in dark matter experiments for gas purification~\cite{lzcharcoal}.  Thus, it was a natural candidate to test for N$_{2}$ radon assays at SNOLAB.   When activated charcoal is used primarily for gas purification, with the goal of preventing any radon atoms from escaping, about a kilogram  of this radon adsorbent is used. For our purpose, to achieve the high radon desorption efficiency  required for screening radon concentration in gas, the amount of charcoal needed is on the order of grams. 

For this paper  Calgon OVC 4 $\times$ 8~\cite{calgon}, which is a granular activated carbon made from coconut shell is used. The charcoal was  purified at SNOLAB and then was used to make a radon trap. The background and radon trapping efficiency of the trap were measured. The latter was performed under various cryogenic temperatures, gas flow rates and flow times.  To measure the radon trapping efficiency, a radon calibration source was developed and characterized.  In addition, the performance of the trap was compared with that of bronze wool using underground lab air and   N$_{2}$ from a cylinder.   The outline of the paper is as follows. the experimental apparatus is detailed in Section~\ref{sec-apparatus}.  The activated charcoal radon assay and analysis methodology, as well as, the radon calibration source characterization are described in Section~\ref{sec-methodology}.  The background and efficiency of the charcoal trap, as well as a comparison of its performance with underground lab air and N$_{2}$	  assays to that of bronze wool, are presented in Section~\ref{sec-results}. In addition, the performance and sensitivity of our radon system  with the addition of the charcoal trap was compared with the original system without the charcoal trap.   The conclusion and discussion are provided  in Section~\ref{sec-discussion}.

 \section {Apparatus}
 \label{sec-apparatus}
 \subsection{Radon board}
\begin{figure}[h]
     \centering
\includegraphics[width=\linewidth]{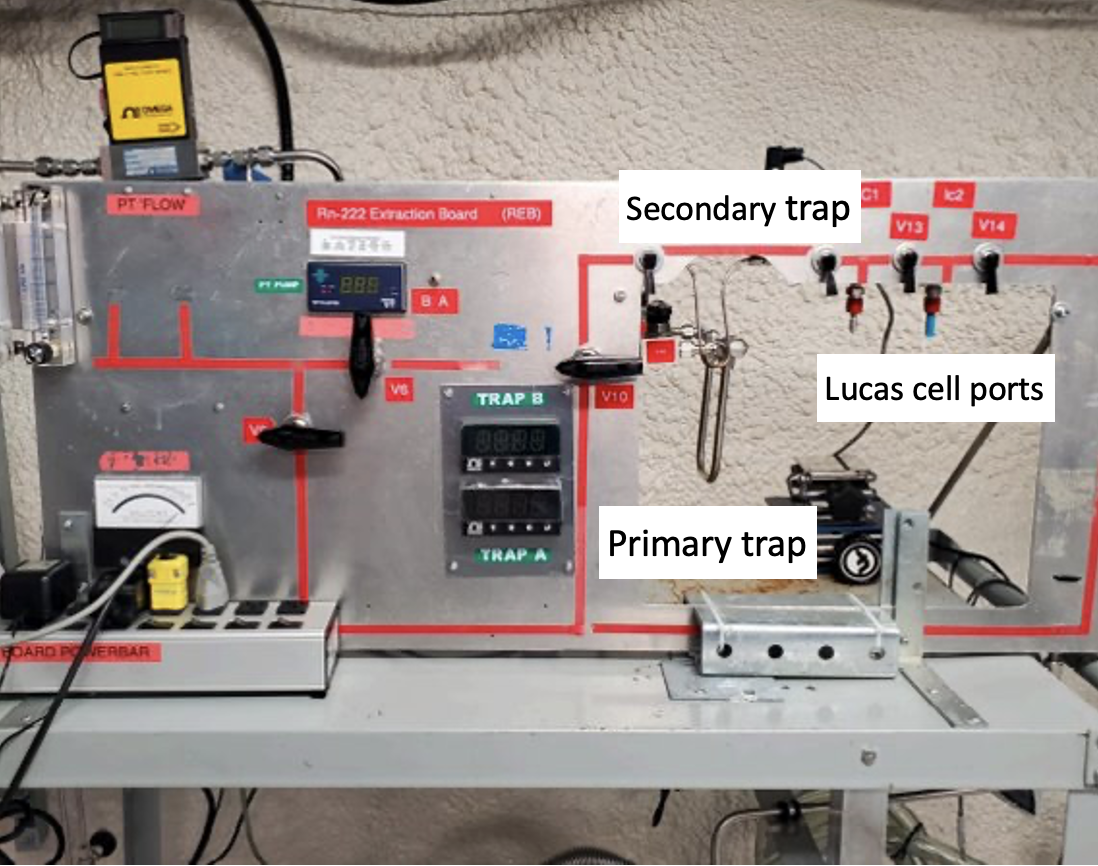}
     \caption{Picture of SNOLAB's underground radon board}
     \label{fig-radonboardpicture}
 \end{figure}

\begin{figure}[h]
    \centering
    \setlength\fboxsep{0pt}
    \setlength\fboxrule{0.25pt}
    \includegraphics[width=0.5\textwidth]{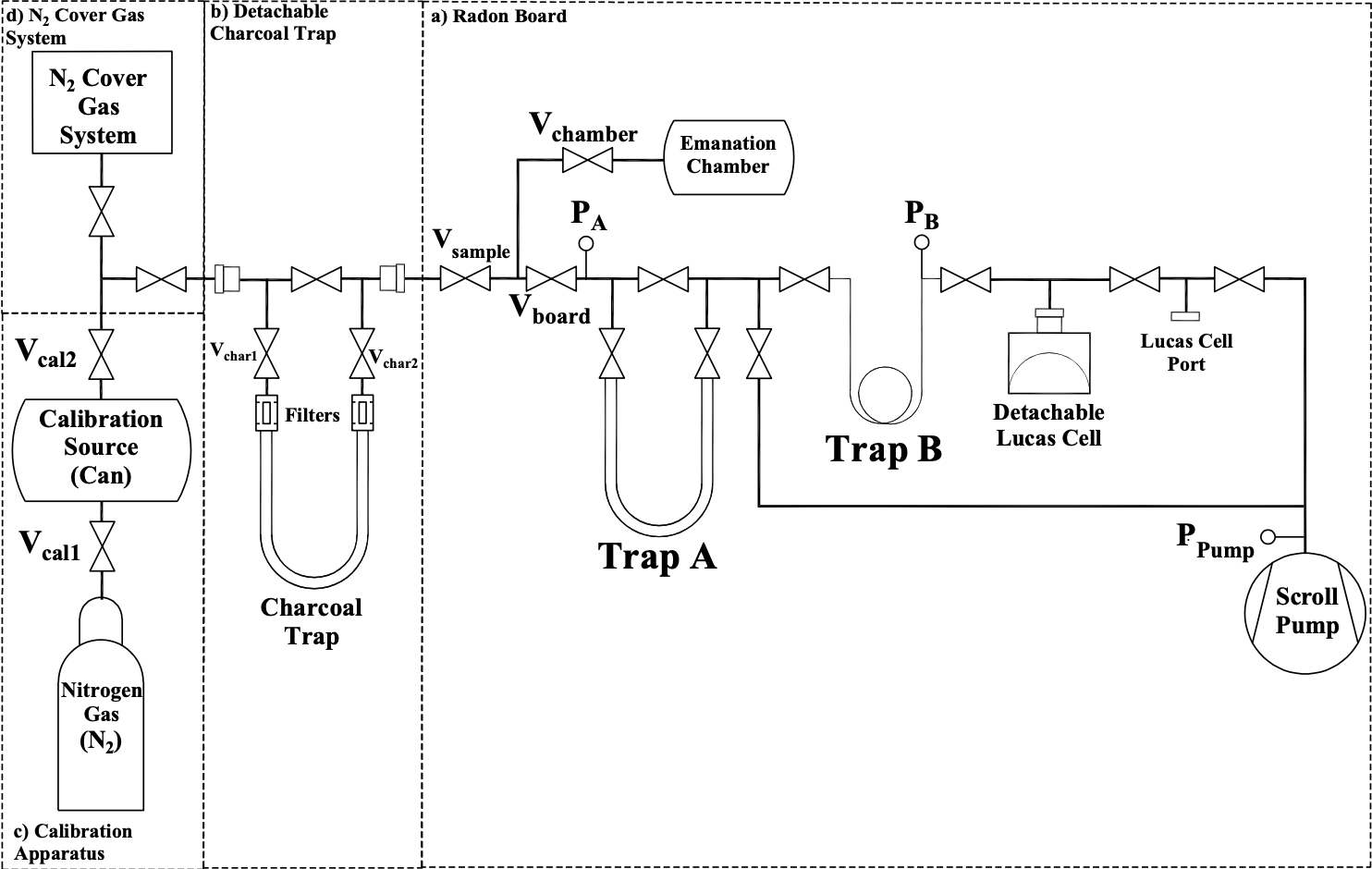}
    \caption{Schematic view of a) a SNOLAB radon board b) portable activated charcoal  trap  c) calibration system apparatus and d) a cover gas system filled with N$_{2}$ }
    \label{fig-radonboard}
\end{figure}

 At SNOLAB, there are two radon boards used for material screening: one is located in the underground lab, and the other is in the surface building (Figure~\ref{fig-radonboardpicture}).  The underground board's tube diameter is 3/8", while the surface board's tube diameter is 1/2". Both boards follow the design described in Ref.~\cite{LIU1993291}. The schematic of these two boards are shown in Figure ~\ref{fig-radonboard}a. They each consisted of   an emanation  chamber, a U-shaped primary radon trap (Trap A), a concentration trap (Trap B), a detachable Lucas cell (LC) and a scroll pump capable of creating a vacuum of  $<5.0 \times 10^{-3}$ mbar. The emanation chamber is used to place  material samples for $^{222}$Rn measurements. Trap A is filled with fine bronze wool to enhance radon trapping.  The boards are equipped with a high-purity mass flow meter and three pressure transducers to monitor gas flow and pressures in both traps, as well as  inlet pressure going into the vacuum pump.
The underground board was built in 1993 from 316L stainless steel and  Swagelok vacuum fittings during the SNO construction and was refurbished in 2019. It is regularly leak checked using a He mass spectrometer for an acceptable leak rate of $<1 \times 10^{-7}$~mbar$\cdot$L/s. The surface board was made from electro-polished 316L stainless steel, with most  components being orbital welded which reduces the frequency of required leak checks for an acceptable rate of $<1 \times 10^{-9}$~mbar$\cdot$L/s. The surface board background is  $0.6\pm 2.9$  %
radon atoms per hour of emanation.   The underground board background is measured to be $1.1\pm 3.0$ radon atoms %
per thirty minutes of emanation. %

 The radon boards can be used for both radon emanation measurement and gas assays. When  used for gas assay the board is connected to a sample gas port through a vacuum fitting, while the valve to the emanation chamber is kept closed.   For N$_{2}$ gas flow of 1~SLPM the efficiency of the bronze wool varies from 85\% to less than 3\%, depending on the duration that gas flows through the trap~\cite{Adilthesis}. This indicates  that a 3/8" diameter tube filled with bronze wool   is  not a good design for performing high sensitivity gas assays. 
\subsubsection{Activated charcoal trap}
A new detachable activated charcoal trap has been developed to perform N$_{2}$ assays with flow rate $\geq~1$~SLPM.  Calgon OVC 4$\times$8 was selected due to  its availability and  low radon emanation~\cite{Xenoncharcoal}.  The $^{238}$U activity of Calgon OVC 4 x 8, before any treatment, was measured to be  $465\pm 47$~mBq/kg from $^{226}$Ra decay using SNOLAB's Canberra Well detector~\cite{IanTAUP}\cite{Wellresults}. Subsequently,  the charcoal was  etched using 30\% diluted nitric acid  with ultra-pure water   for 18 hours, followed by rinsing with ultra-pure water.  After etching and rinsing, the charcoal was  baked at 150~$\degree$C while it was under vacuum pressure of $1.0 \times 10^{-3}$~mbar for 72 hours to ensure thorough drying.   %
The $^{238}$U activity of cleaned Calgon charcoal was measured to be $<33$ mBq/kg at 90\% CL from $^{226}$Ra decay by the SNOLAB Well detector~\cite{Wellresults}. Thus, the nitric acid washing procedure described above reduced $^{226}$Ra by one order of magnitude. 

\begin{figure}
    \centering
\includegraphics[width=0.7\linewidth]{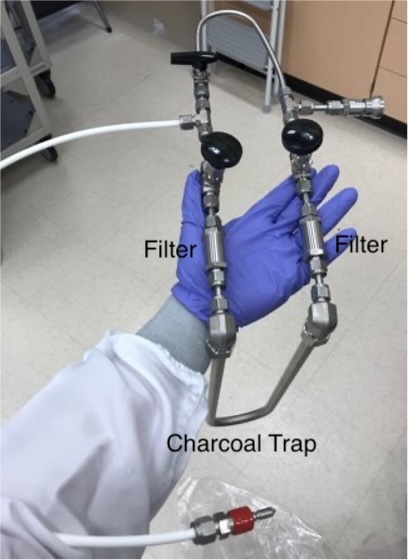}
    \caption{Activated Charcoal trap}
    \label{fig-pictureofu}
\end{figure}

A 1/2 inch diameter, 13-inch length U-tube was filled with 22~g of acid-etched activated charcoal. Two inline  sintered stainless steel filters, with a nominal pore size of $\mathrm{0.5~\mu m}$ from Swagelok~\cite{filters},    were used to prevent charcoal particulates  from escaping the trap. The charcoal trap is shown in Figure ~\ref{fig-pictureofu}.   This portable trap can be connected to both surface and underground radon boards. The surface board was used to characterize the efficiency and background of this trap (Section~\ref{sec-results}) and the underground board is used for performing radon assay of N$_{2}$ cover gas systems and mine air. A schematic of portable charcoal trap while attached to the board through V$_{sample}$ is shown in Figure~\ref{fig-radonboard}b.  %
\subsection{Lucas cell counting system}
\begin{figure}
    \centering
    \includegraphics[width=1.0\linewidth]{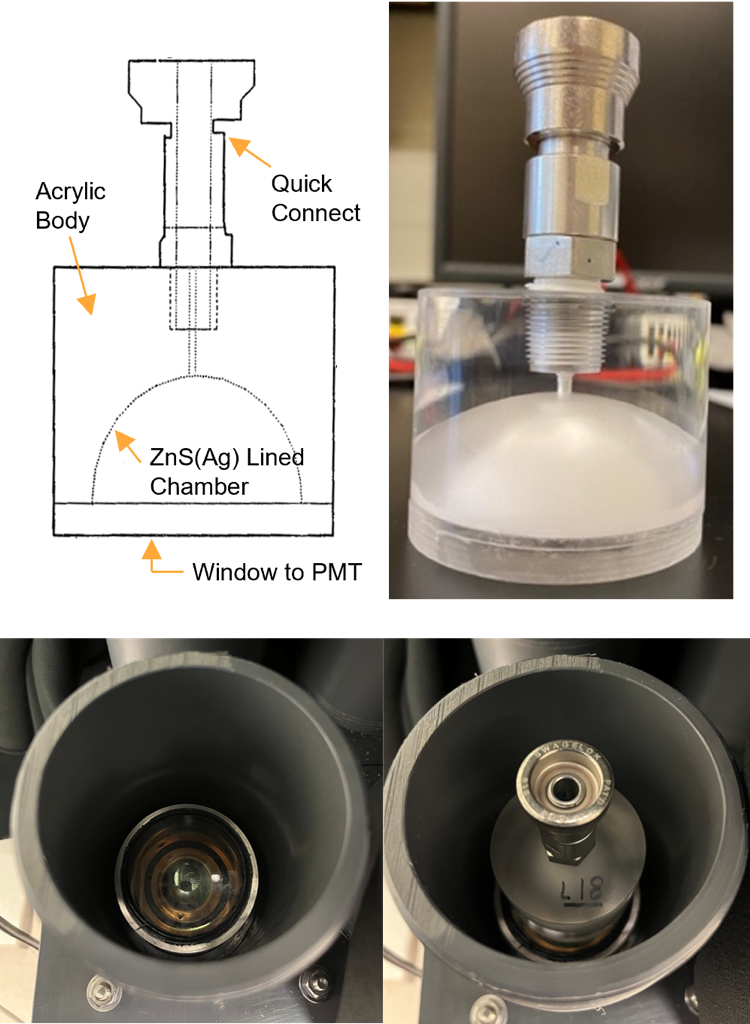}
    \caption{Picture of a Lucas Cell (LC) and one of the  PMTs used for counting.}
    \label{fig-lc}
\end{figure}

A Lucas cell  is a 2-inch scintillation device consisting of a 10 mg/cm$^2$ of silver-activated zinc sulfide, ZnS(Ag), on its inner hemispherical acrylic dome surface. The design of the LC is described in Ref.~\cite{LIU1993291}. A picture of an LC is shown in Figure~\ref{fig-lc}. The efficiency of the LC for detecting   alpha-particles from $^{222}$Rn progeny was measured by the  SNO collaboration to be $74\pm 7\%$~\cite{waterpaper}.
After the final transfer of $^{222}$Rn into the LC, the cell is placed on a photo-multiplier tube (PMT).  Eight HZC PHOTONICS XP2262 PMTs are  used for counting alpha pulses from radon assays. The signal from  the PMTs  are processed by a CAEN N6725 digitiser.  The output of the digitiser is recorded using a  a CAEN  readout application~\cite{wavedump}.
\subsection{Radon calibration source}
\begin{figure}
    \centering
    \includegraphics[width=1.0\linewidth]{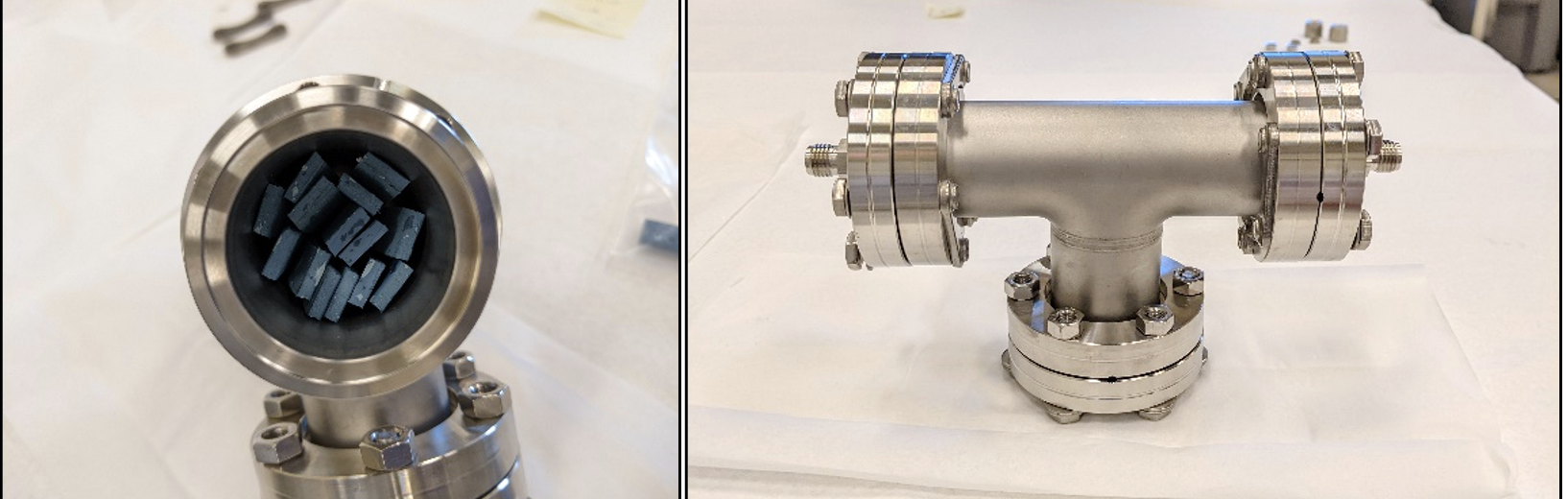}
    \caption{The Calibration source used for the charcoal radon trap efficiency measurement before  and after  closing the cap}
    \label{fig-calibrationcanpicture}
\end{figure}

To measure the efficiency of the radon traps a radon calibration source was made from Nora Xp 5319 rubber floor tile~\cite{floortile}. %
We used  $69\pm 2~\mathrm{g}$ of this  material. It was  cut into fourteen strips of $\mathrm{1~cm\times 10~cm}$ and was inserted in a 2.75 inch ConFlat (CF) tee. The picture of the can before and after closing the cap of the CF flange is shown in Figure~\ref{fig-calibrationcanpicture}.  The calibration source was later characterized using the SNOLAB surface radon board. The radon emanation results of this source are presented in Section~\ref{sec-calibrationcan}. The apparatus to connect the calibration source to the radon board is shown in Figure~\ref{fig-radonboard}c.
\section{Methodology}
\label{sec-methodology}
The procedures followed for each type of $^{222}$Rn assay are slightly dissimilar to one another and are briefly described in their relevant sections with the results. Generally, before and after each assay, the assay board is purged by passing high-purity nitrogen gas through it and then pumped with a vacuum pump to a pressure of order $10^{-3}$ mbar while the traps are heated. A $^{222}$Rn assay begins by drawing a sample gas into the board using the vacuum pump, passing it through Trap~A while it is cooled by submerging it in LN --- this is the first transfer. The cryogenic temperature of the primary trap during this transfer allows radon to lose energy as it passes through it, thereby increasing the surface area with the bronze wool inside the trap. If the sample gas is drawn from the emanation chamber with a sample material inside, the volume of the gas is small enough for the flow rate to not be of significant concern ($<$0.1~SLPM). If the sample gas is drawn from a larger volume, such as the N$_2$ cover gas systems underground or an N$_2$ cylinder, the inlet valve is adjusted to maintain a desirable flow.\\

\noindent After the completion of the first transfer, the trapped gas in the primary trap is transferred to Trap B (with a factor of ten smaller volume) by first isolating the two traps and the path between them, followed by the heating of the primary trap to 100$^{\circ}$C while the Trap B is cooled by LN. This second transfer occurs through the process of cryo-pumping whereby radon, and other trapped gases in the Trap A are released  by heating and settle in the cryo-cooled concentrator trap. Cryo-pumping is a highly efficient process for low-pressure gases within a small transfer path length and volume~\cite{cryopumping}. Transferring to a smaller volume trap improves the efficiency of the following final transfer between Trap B and an LC by volume sharing while Trap B is heated to $100~\degree$C. %
Upon completion, the LC is detached from the assay board and placed on top of a PMT  to be counted.\\
\subsection{Activated charcoal trap testing}
The efficiency of $^{222}$Rn adsorption and desorption of the activated charcoal trap is measured across a range of flow rates from  1 SLPM to 7 SLPM. A combination of $^{222}$Rn emanation from the test source (calibration can) and N$_2$ as a carrier gas is used to mimic underground cover gas systems. In this case, the first transfer is through the activated charcoal trap using the vacuum pump while the trap is cooled. Initially, LN was used in order to cool down the trap, but it was discovered that LN slows the N$_{2}$ molecules, resulting in a high pressure within the trap ($>$30 PSI)  and subsequently increased uncertainty on the assay results. To overcome this issue, alcohol slush --- a mixture of reagent alcohol and LN --- was used as the  cryo-liquid. The temperature of the cryo-liquid was varied in order to measure the efficiency of trap versus temperature.    The temperature of the external wall  of the U-tube was measured using a dedicated sensor.  The subsequent transfers are the same as described  earlier (Charcoal $\rightarrow$ Trap A $\rightarrow$ Trap B $\rightarrow$ LC).\\ 
\subsection{Data  processing and analysis}
  The subsequent alpha decays from $^{222}$Rn, $^{218}$Po and $^{214}$Po allows the determination of  the number of radon atoms that decayed  in the LC.  The alpha decay from $^{210}$Po is a source of background. As $^{210}$Pb has a halflife of 22 years,  $^{210}$Po background increases for LCs which are used regularly in the assays. Other sources of alpha background from natural radioactivity remain constant and are measured to be $3\pm 1$ counts per day, assessed just after fabricating the cells.  For measuring the background of the radon boards and radon traps, LCs with background of $3\pm 1$ to $5\pm 1$ counts/day were used. For measuring the transfer and trapping efficiencies, where a source with high radon emanation rate was used, LCs which had backgrounds of $20\pm 2$ to $100 \pm 3$ counts/day were used.  
  
  To measure number of radon atoms in a given source, first the number of $\alpha$ decays are counted and then were converted to radon atoms from a radon source,
\begin{equation}
\label{eq-alphatoradon}
    N_{Rn}=\frac{N_{\mathrm {count}}-N_{LCB}}{3\epsilon_{LC}\epsilon_{R}(1-e^{-\lambda t_{\mathrm{count}}})}-R_{B},
\end{equation}
  where $N_{\mathrm{count}}$ is the cumulative number of detected $\alpha$ counts, $t_{count}$ is the total counting time, $N_{LCB}$ is the number of LC background for the counting period of $t_{count}$, $\epsilon_{LC}$ is the LC alpha efficiency (which is multiplied by 3 for the correlated alpha decays from $^{222}$Rn progeny), $\epsilon_{R}$ is the total  efficiency of radon board in transferring radon atoms from  the primary trap to LC,  $\lambda$ is the decay constant for $^{222}$Rn, $R_{B}$ is the background of the radon board, including the traps during the assay.   The total assay time for all the results presented in this paper was less than or equal to two hours. The number of radon atoms that decay during a two hour measurement is considered to be negligible compared to total systematic and statistical uncertainties on the measurement and thus not included in the equation. We used $74\pm 7$\% for the LC efficiency, as known from Ref.~\cite{waterpaper}. The radon board efficiency was measured using a procedure described in Section~\ref{sec-eff}. In addition to statistical uncertainty on $N_{count}$, we included a systematic uncertainly on the counting introduced by the PMT and the electronics channels. We counted one high background cell with a constant amount of alpha decays from $^{210}$Po using the eight PMT channels used for counting and we found the counted number of alphas differed by 4.2\% across the eight channels.

\subsection{Radon board efficiency measurement}
\label{sec-eff}
The efficiency of the radon board is the combination of  Trap A efficiency,  cryo-pumping transfer efficiency from Trap A to Trap B, and transfer efficiency from Trap B to the LC.  As the transfer of radon atoms from Trap B to LC is done through volume sharing, the latter can be calculated by looking at the value of the pressure transducer (P$_{\mathrm{B}}$) before and after opening the valve between the LC and Trap B.  The transfer efficiency from  Trap B to LC is calculated to be $83\pm 1$\% and $81\pm 1$\% for the surface and UG boards, respectively. The uncertainty on this measurement was found by performing five measurements for each board. To measure combined transfer efficiency, an LC   was filled with radon atoms and then counted using the counting system to determine the exact amount of radon atoms in the cell. The LC was then attached to V$_{sample}$ via a quick connect. While Trap A was at LN temperature and under vacuum, $V_{sample}$ was opened and the LC was pumped through Trap A for an hour, following a  procedure similar to that used for performing radon assays. By the end of the  hour long transfer, the LC pressure reached the ultimate pressure of the vacuum pump ($5.0\times 10^{-3}$~mbar). The rest of the procedure was followed to direct the radon atoms to  Trap B. In order to remove any systematic uncertainty related to LC efficiency,  the LC was detached from V$_{sample}$ and was attached to the LC port to transfer radon atoms from Trap B to the same cell.  The total efficiency of the board is  defined as:
\begin{equation}
 \epsilon_{R}=   \frac{N^{tr}_{LC}}{N^{in}_{LC}},
\end{equation}
where $N^{tr}_{LC}$ is number of radon atoms transferred and collected in the cell after applying the above procedure and $N^{in}_{LC}$ is number of radon atoms initially present in the cell before the transfer.  The surface board total efficiency was found to be $80\pm 6$\% and the UG board total efficiency was found to be $75 \pm 6$\%. The uncertainties  on these measurements are from repeating the same procedure five times for each of the boards and calculating the standard deviation of the five measurements.    The combined Trap A and Trap B efficiency is defined as:
\begin{equation}
\epsilon_{trap}=\frac{\epsilon_{R}}{\epsilon_{vol}},
\end{equation}
where $\epsilon_{vol}$ is  Trap B to LC radon transfer efficiency due to volume sharing. This gives the combination of Trap A and Trap B  efficiency in trapping radon atoms to be $96^{+4}_{-8}\%$ and  $94^{+6}_{-7}\%$ for surface and underground board, respectively.   
\subsection{Systematic uncertainties}
\label{sec-uncertainty}
All the terms that enter Equation~\ref{eq-alphatoradon} and their systematic uncertainties are summarized  in Table~\ref{table-uncertainties}. While knowing the $N_{LCB}$  uncertainty is important for background measurement of the radon board and charcoal trap, it is a negligible source of uncertainty for assays using the calibration can and underground lab air. The  uncertainties are combined in quadrature to determine the overall systematic uncertainty of $14\%$.
\begin{table}[]
    \centering
    \begin{tabular}{c|c|c}
    Variable &  Value & uncertainty \\
    \hline
     $N_{count}$ &  $N_{count}$ & $N_{count} \times 0.04$ \\ 
    $N_{LCB}$   &  $5$ & $1$ \\
                &  $3$ & $1$ \\  

    $\epsilon_{LC}$ 
         &  $74\%$  & $7\%$ \\
    Surface board, $\epsilon_{R}$   & $80\%$  &  $6\%$\\
    Underground board, $\epsilon_{R}$ & $75\%$  & $7\%$ \\
    Boards background & 0  & \\ 
    \end{tabular}
    \caption{Terms used to determine the number of radon atoms and the overall systematic uncertainty in the assays.}
    \label{table-uncertainties}
\end{table}

\subsection{Radon source characterization}
\label{sec-calibrationcan}
\begin{figure}
    \centering
    \includegraphics[width=\linewidth]{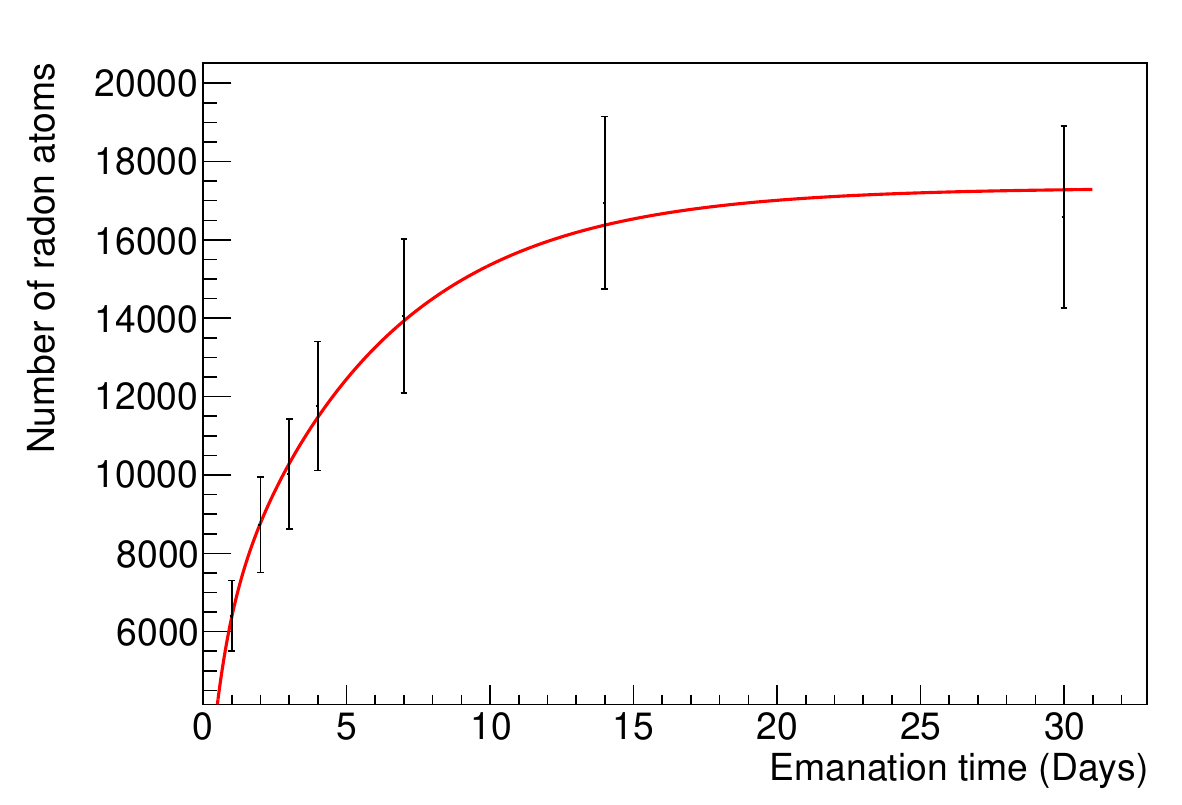}
    \caption{Number of atoms emanated in the  can used for charcoal trap calibration versus  the emanation time. }
    \label{fig-calibrationcan}
\end{figure}

The calibration source used for charcoal trap efficiency measurement  was emanated under vacuum and was  characterized.  For this work, the can was connected to the board through $V_{sample}$, while Trap A was under LN and the valve between the can and trap A was open.  A flow of 0.1 SLPM of N$_{2}$ went through the calibration can for an hour while Trap A was cooled with LN. This small flow rate was used   to  help with radon transfer from the calibration can to Trap A. The number of radon atoms emanated in the can, $N_{Rn}$, was measured for various emanation times. The results are shown in Figure~\ref{fig-calibrationcan}. The statistical and systematic uncertainties are shown using a single combined error bar.  In order to predict the number of radon atoms in the calibration can for any emanation time not directly measured,  a mathematical model was developed and fitted to the available data. We assumed that radon emanated from the rubber floor tiles is supported by $^{226}$Ra decay, as well as outgassing of $^{222}$Rn atoms that are trapped within the material's pores. The number of $^{222}$Rn atoms in the calibration can  be defined as:
\begin{equation}
N^{\mathrm{can}}_{Rn}(t)=N_{Ra}(t)+ N^{\mathrm{can}}_{\mathrm{outgassing}}(t),
\end{equation}
The first term represents the number of radon atoms produced  by radium decay and  can be written as:
\begin{equation}
N_{Ra}(t)=p_{0}(1-e^{-\lambda t}),
\end{equation}
where $\lambda$ is the decay constant of $^{222}$Rn.
The second term represents the number of radon atoms resulting from the outgassing process, which have left the material's surface and entered the volume of the can. Assuming that the probability of radon outgassing from the material is proportional to the number of radon atoms trapped in its pores, we can write:
\begin{equation}
\begin{split}
    \frac{dN_{\mathrm{pores}}}{dt} &=-(\lambda^\prime+\lambda) N_{pores}, \\
\frac{dN^{\mathrm{can}}_{\mathrm{outgassing}}}{dt} &=\lambda^\prime N_{\mathrm{pores}}-\lambda N^{\mathrm{can}}_{\mathrm{outgassing}},
\end{split}
\end{equation}
where $N_{\mathrm {pores}}$  represents the number of radon atoms trapped inside the pores, and  $\lambda^\prime$ is the probability of  outgassing for the trapped radon atoms.  By solving these two differential equations, we obtained:
\begin{equation}
   N^{\mathrm{can}}_{\mathrm{outgassing}} (t) = p_{1}e^{-\lambda t}(1-e^{-\lambda^\prime t}),
\end{equation}
where $p_{1}$ represents  the total radon atoms trapped inside the pores at $t=0$.
The final mathematical model fitted  to the data points is:
\begin{equation}
   N^{\mathrm{can}}_{Rn}(t)=p_{0}(1-e^{-\lambda t})+p_{1}e^{-\lambda t}(1-e^{-\lambda^\prime t}).
\end{equation}
The fit function is shown in Figure~\ref{fig-calibrationcan} using red line.  
The method of $\chi^{2}$ was used to perform the fit.  $p_{0}$,   $p_{1}$ and $\lambda^\prime$ were free parameters in the fit.   These fitted parameters were used to estimate number of radon atoms in the calibration can for the emanation time periods for which  we did not have data points. 

\section{Results}
\label{sec-results}
\subsection{Background measurement of the charcoal trap}
Two sets of measurements were performed in order to determine the charcoal trap $^{222}$Rn background. For the first set, the charcoal trap was purged with N$_{2}$ and was baked for 20 minutes. It was  left to be emanated under vacuum at room temperature  for 4 and 17 days. Then the charcoal trap was connected to $V_{sample}$ and emanated radon atoms from the charcoal were transferred to Trap A using the vacuum pump for an hour. The charcoal trap was heated to $150~\degree$C during the transfer, to increase the thermal energy of the radon atoms to break the Van der Waals bonds with charcoal molecules.  The transfer of radon atoms to Trap B and an LC was performed using the same procedure described in the methodology. The two emanation rate measurements were $36 \pm 6$ radon atoms/day and $34\pm 6$ radon atoms/day.  The error bars consist of a 14\% systematic uncertainty, as well as the statistical uncertainty, which was added in quadrature.    

The second set of measurements was performed in order to determine the number of radon atoms in charcoal in the same operating conditions under which  gas assays are performed.  For normal assays this trap is used in total for two hours. During the first hour the trap is under vacuum and at  a cryogenic temperature of -70~$\degree$C.  In the second hour, the trap is warmed up to $150~\degree$C . For this work a cleaning procedure was developed  to remove any residual  radon atoms  from previous assays. Initially, the trap was purged with N$_{2}$ from a cylinder, but later it was discovered that N$_{2}$ purging elevates the background of the trap, indicating that N$_{2}$ from a cylinder is itself is a source of radon. N$_{2}$ purging was therefore removed from the cleaning procedure.  It was also determined that in order to remove all residual radon atoms remaining in charcoal pores from previous assays the trap needs to be baked for at least two hours at $150~\degree$C.  The background of the charcoal trap was then measured as it was under vacuum and was cooled  to $-70\degree$~C for one hour,  then later heated at 150~$\degree$C for one hour. This is similar to the operating conditions of this trap. This measurement was performed twice and the number of radon atoms was found to be $4.6\pm 2.8$, using the average value of the two measurements.  This background level is used for the all charcoal trap assays reported in this paper. SNOLAB's Well detector set the limit of  $<$~33 mBq/kg~\cite{Wellresults}, corresponding to $<$~2.6 radon atoms per hour of assay time for the 22~g charcoal which is in the U-trap. The results between two methods are less than two sigma away from each other. For the Well detector's measurement, 3.2~g of charcoal was used in comparison with 22~g used in the trap. Therefore the results are consistent statistically.  

\subsection{Charcoal trap efficiency to N$_{2}$ gas assay}
To find the optimal operational condition of the activated charcoal trap which gives highest trapping efficiency, some conditions were varied. These included the temperature of the cryo-liquid, the assay duration, and the  mass flow rate. 
\subsubsection{Efficiency versus temperature}
\begin{figure}
    \centering
    \includegraphics[width=1\linewidth]{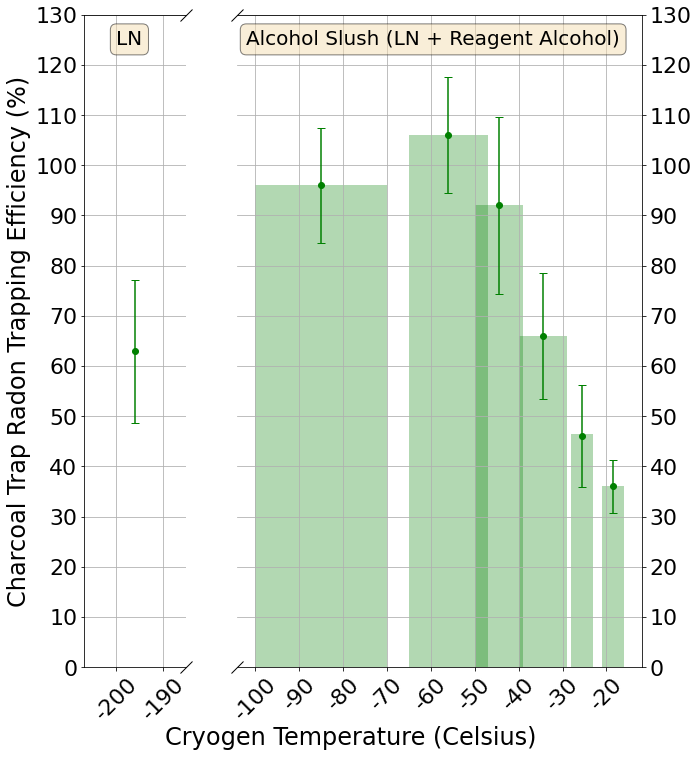}
    \caption{Efficiency of the  activated charcoal trap versus cryogenic temperature for flow rate of 5 SLPM and assay duration of one hour. }
    \label{fig-tempraturevariation}
\end{figure}

Charcoal trap efficiency was measured  from a temperature of -196~$\degree$C (LN temperature) to -20$\degree$C for a mass flow rate of 5~SLPM and one hour of assay time. The results are shown in Figure~\ref{fig-tempraturevariation}.  The variation of the temperature for each data points (green bins)  is due to the increasing of temperature of the alcohol slush during the  one hour of assay.  This was due to the fact that the alcohol slush was made before starting the assays  and the top of the coolant used was exposed to room temperature during the procedure.  The overlapping bins in the graph is due to the overlapping temperature of alcohol slush. Therefore the measurements in the overlapping regions are not statistically independent.  From this graph it was concluded that the alcohol slush temperature of $<$$-50~\degree$C yields a 100\% trapping efficiency.  It was decided that the trap should be operated at a starting alcohol slush temperature of $-80~\degree$C and the temperature of the alcohol slush is monitored during an assay to ensure the temperature is maintained to be $<-50~\degree$C.

\subsubsection{Efficiency versus mass flow rate and assay duration}
Tests were performed using a one hour assay duration for flow rate of 1, 3, 5 and 7~SLPM while the slush temperature was kept between $-80~\degree$C to $-65~\degree$C. It was found that the relative efficiency of the charcoal trap stayed constant for these flows and was consistent with 100\%. 

Maintaining the temperature of the alcohol slush in the region $<$$-50~\degree$C for an assay duration of longer than two hours was challenging and therefore it was decided to set  the maximum assay time to two hours.  It was found out that for a flow rate of 5 L/min the efficiency of charcoal trap for a two hour assay was  $(87.5\pm 12.0)\%$.  

\subsection{Comparison of the efficiency of charcoal trap with bronze wool trap}
\begin{figure}
    \centering
    \includegraphics[width=\linewidth]{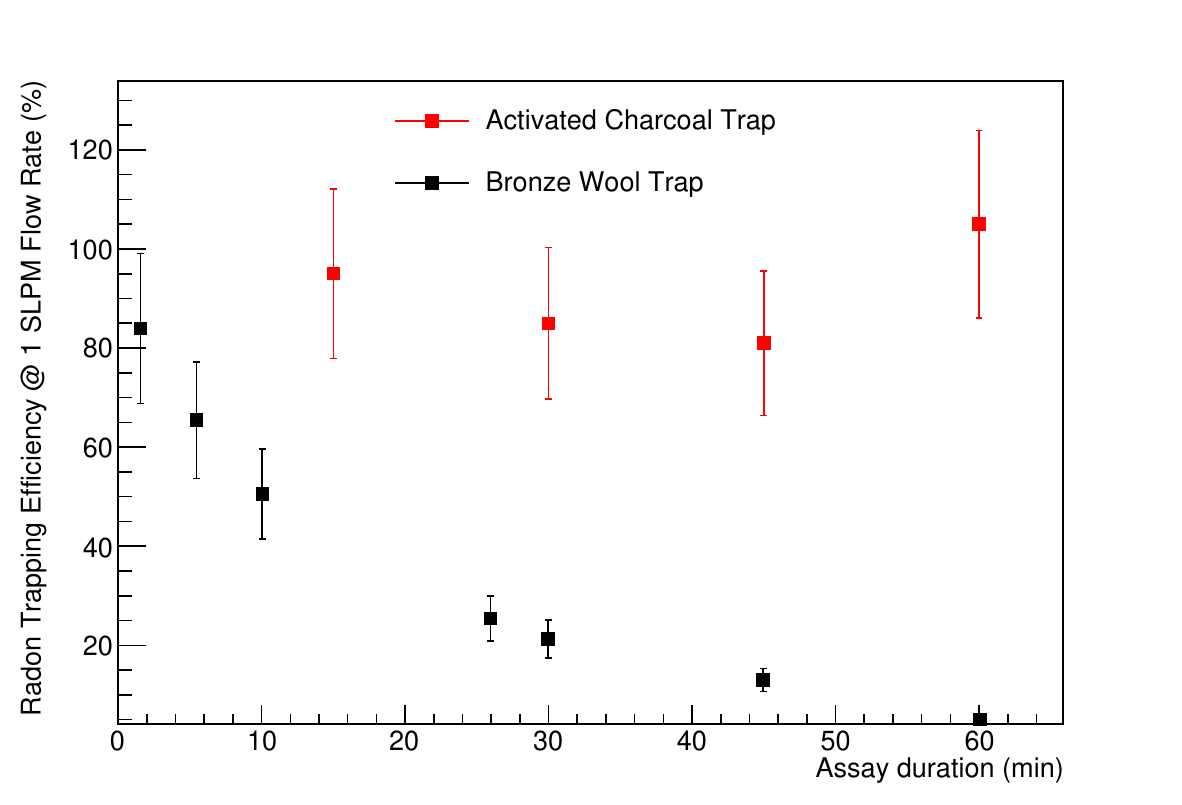}
    \caption{Comparison of bronze wool efficiency versus charcoal trap efficiency at flow rate of 1~SLPM.}
    \label{fig-mineair}
\end{figure}

The efficiency of the bronze wool trap on the underground board at 1 SLPM was measured using lab air, as described in Ref. \cite{Adilthesis}. The lab air passed through a drierite and a sodium hydroxide column to remove water moisture and CO$_2$, respectively. A RAD7~\cite{RAD7}, situated next to the radon board, was used to measure the $^{222}$Rn activity in the lab air in units of Bq/m$^3$. The $^{222}$Rn activity was converted to the number of radon atoms for a given volume of lab air using the $^{222}$Rn decay constant. The efficiency of the bronze wool trap was defined as:
\begin{equation}
\epsilon_{bronze}= \frac{N_{counts}}{\epsilon_{LC}\epsilon_{\mathrm{trapB}}N_{\mathrm{RAD7}}},
\end{equation}
where $N_{\mathrm{RAD7}}$ is the number of radon atoms calculated using the activity measured by the RAD7. The efficiency of the bronze wool trap at 1 SLPM versus various assay times is shown with black points and error bars in Figure~\ref{fig-mineair}. In addition to the 14\% systematic uncertainty described earlier, there is an additional 10\% uncertainty in the measurement due to fluctuations in the flow rate and a 5\% uncertainty in the RAD7 radon measurement. The RAD7 measures radon concentration in the lab every two hours, with 12 measurements per day. The standard deviation of the 12 measurements taken on the same day as the assays was used to estimate the 5\% uncertainty in the RAD7 measurement. The systematic uncertainties were added in quadrature. For comparison, the efficiency of the charcoal trap at the same flow rate of 1 SLPM for various assay times is shown in red. The charcoal trap temperature was between -70~$\degree$C to -45~$\degree$C for all the measurements.  Unlike the bronze wool, where the efficiency decreases rapidly with assay time, the charcoal trap maintains a consistent efficiency for assay times ranging from 15 to 60 minutes.

Furthermore, we measured the efficiency of the bronze wool trap at a flow rate of 5 SLPM after one hour of assay time using the calibration can. The trap was cooled with liquid nitrogen (LN). We observed alpha rates consistent with the LC background, indicating that the efficiency of the bronze wool trap is effectively zero at 5~SLPM.
\subsection{N$_{2}$ gas assay sensitivity using charcoal trap}
The sensitivity of the radon board at 95\%~CL, with the addition of the charcoal trap, to radon per cubic meter of N$_{2}$ was determined using the following equation:
\begin{equation}
    S=\frac{R+3\sigma_{R}}{V.\epsilon},
\end{equation}
where $R$ is the background of the radon board and the charcoal trap for the duration of assay, $\sigma_{R}$ is the uncertainty on the background value, $V$ is the volume of gas extracted and $\epsilon$ is the efficiency of the charcoal trap which is considered to be 100\% at temperature of -60~$\degree$C. For an assay time of 1 hour with a flow rate of 5~SLPM, 0.3~m$^{3}$ of gas is extracted. The sensitivity of the radon board with the charcoal trap installed is $>$~$9.1\times 10^{-2}$~mBq/m$^3$ at 95\%~CL. The underground board, using only bronze wool as the primary trapping mechanism, operates at 1~SLPM for an assay time of 30~minutes, allowing for 0.03~m$^{3}$ of gas extraction with a radon trapping efficiency of 21\%. Under these operational conditions, the sensitivity of the radon board with the bronze trap is $>$3.3~mBq/m$^3$ at 95\%~CL. The introduction of the charcoal trap to the radon board improves the board's sensitivity by at least a factor of thirty.

\subsection{Measuring radon level in $N_{2}$ cylinder using charcoal and bronze trap}
We measured the concentration of radon in a Grade 5 N$_{2}$ cylinder~\cite{n2} using the bronze wool and activated charcoal trap, following the procedures described above. For the bronze wool measurement, we performed a 40~L gas assay at flow rate of 1~SLPM.    For the charcoal trap measurement, we extracted 300~L at a  flow rate of 5~SLPM. The results obtained with the bronze trap were consistent with LC background and corresponded to $2.3\pm 4.2$ Rn atoms/L of $N_{2}$.  The results using charcoal trap was  $0.6 \pm 0.1$ Rn atoms/L of N$_{2}$.
 N$_2$ assay results confirms that we have improved our capability at SNOLAB to measure low concentration of $^{222}$Rn in N$_2$.

\section{Conclusion and discussion}
\label{sec-discussion}
In summary, the purpose of this work was to improve sensitivity of $^{222}$Rn measurement in the N$_{2}$ cover gas systems at SNOLAB. We built a radon trap filled with nitric acid washed Calgon OVC $4\times 8$ activated charcoal. We tested the trap at various flow rates, assay times and temperatures to  determine the optimal operational temperature in order to have maximum efficiency.  The trap has sensitivity of  $>$~$9.1\times 10^{-2}$~mBq/m$^3$~@ 95\%~CL for gas flow rate of 5~SLPM and and assay duration of one hour.   To improve this sensitivity, we need to reduce the trap background of ($4.6\pm 2.8$) atoms/h  and to be able to extract gas at higher flow rate. We plan to try different nitric acid concentration in order to see if we can reduce the background of the charcoal trap further. We also observed that the 0.5-micron filters we used restrict the flow by at least a factor of 5. These filters are 3~cm long, and in the future, we plan to test shorter filters to reduce flow restriction.

We anticipate that some of the systematic uncertainty in the trapping efficiency arises from the uncertainty on the trap's temperature during the cooling and warming up process. For future improvements, we aim to develop a more robust method for cooling and warming  the trap, including replacing the alcohol slush with immersion coolers.  Additionally, we plan to conduct research to use the trap for measuring radon concentrations in different noble gases.

\section*{Acknowledgments}
We would like to thank Julia Azzi, Madeleine Berube, and Keegan Paleshi for their contributions to this work during their co-operative placements. We appreciate Stephen Sekula, Alex Wright and Steffon Luoma for proofreading this article. We are thankful to Lina Anselmo, Jeter Hall, Mark Ward, Christine Kraus, and Aleksandra Bialek for their helpful discussions about this research. We are grateful to Deena Fabris and Sharayah Read for performing the acid purification of the activated charcoal, and to Ian Lawson  for conducting the Well Detector measurements. We would like to thank SNOLAB and its staff for support through underground space, logistical and technical services. SNOLAB operations are supported by the Canada Foundation for Innovation and the Province of Ontario, with underground access provided by Vale Canada Limited at the Creighton mine site.


\begin{thebibliography}{10}

\bibitem{snoplus}
SNO+ Collaboration.
\newblock {\em JINST}, 16 P08059, 2021
\bibitem{deap3600}
DEAP-3600 Collaboration. 
arXiv: 1712.01982, 2018
\bibitem{ug-radonlevel}
I. Lawson. 
\newblock{\em TAUP2023 proceeding}, https://doi.org/10.22323/1.441.0303
\bibitem{LN-plant}
Blaire Flynn et al, 
\newblock {\em Cryogenic Society of America, inc}, Cold Fact, September 2023 
\bibitem{n2}
NITROGEN, N2, SMARTOP, ALPHAGAZ 1, 50, (8.1M3 /286.05SCF), CGA-580
%
%
\bibitem {LIU1993291}
M. Liu et al,
\newblock{\em NIM. A Vol. 329}, issue 1-2 p 291:298, 1993
\bibitem{waterpaper}
I. Blevis et al,
\newblock{\em NIM. A Vol. 517}, 139:153, 2004
\bibitem{Adilthesis}
S. A. Hussain, 
\newblock{\em MSc thesis}, 
\url{https://zone.biblio.laurentian.ca/handle/10219/3867}, Accessed date: January  2025
\bibitem{heatcapacity}
The Engineering ToolBox, \url{https://www.engineeringtoolbox.com/specific-heat-solids-d_154.html}, Accessed date, January  2024
\bibitem{lzcharcoal}
The LUX-ZEPLIN (LZ) Collaboration,
\newblock{\em Eur. Phys. J. C }, 80:1044, 2021
\bibitem{calgon}
Calgon Carbon, A Kuraray Company, \url{https://www.calgoncarbon.com/app/uploads/DS-OVC4x815-EIN-E2.pdf}, Accessed date: January  2025 

\bibitem{Xenoncharcoal}
K. Pushkin et al, 
\newblock{\em NIM. A Vol. 903},  267:276, 2018
\bibitem{IanTAUP}
I. Lawson,
\newblock{\em PoS, Vol. 441}, TAUP-2023 proceeding, 2023
\bibitem{Wellresults}
SNOLAB low background page, \url{https://www.snolab.ca/users/services/gamma-assay/Canberra_Well_snolab_master.html}, Accessed date: January  2025
\bibitem{filters}
Swagelok product, \url{https://products.swagelok.com/en/c/inline-filters/p/SS-4F-05}, Accessed date: January 2025
\bibitem{wavedump}
CAEN Digitizer readout application, \url{https://www.caen.it/products/caen-wavedump/}, Accessed date: January 2025
\bibitem{floortile}
Nora floor tile, \url{https://www.nora.com/canada/en/products/norament-xp-trac}, Accessed date: January 2025
\bibitem{cryopumping}
C. Day, 
\newblock{Handbook of Surfaces and Interfaces of Materials, Vol 5}, 265:307, 2001

\bibitem{RAD7}
DURRIDGE RAD7 Radon Detector, \url{https://durridge.com/products/rad7-radon-detector/}, Accessed date: January 2025

%
%

%


\end{thebibliography}
\end{document}